\documentstyle[twoside]{article}

\catcode`\@=11
\long\def\@makefntext#1{
\protect\noindent \hbox to 3.2pt {\hskip-.9pt  
$^{{\eightrm\@thefnmark}}$\hfil}#1\hfill}               

\def\@makefnmark{\hbox to 0pt{$^{\@thefnmark}$\hss}}    
        
\def\ps@myheadings{\let\@mkboth\@gobbletwo
\def\@oddhead{\hbox{}
\rightmark\hfil\eightrm\thepage}   
\def\@oddfoot{}\def\@evenhead{\eightrm\thepage\hfil
\leftmark\hbox{}}\def\@evenfoot{}
\def\sectionmark##1{}\def\subsectionmark##1{}}

\oddsidemargin=\evensidemargin
\newcounter{sectionc}\newcounter{subsectionc}\newcounter{subsubsectionc}
\renewcommand{\section}[1] {\vspace{12pt}\addtocounter{sectionc}{1} 
\setcounter{subsectionc}{0}\setcounter{subsubsectionc}{0}\noindent 
        {\tenbf\thesectionc. #1}\par\vspace{5pt}}
\renewcommand{\subsection}[1] {\vspace{12pt}\addtocounter{subsectionc}{1} 
        \setcounter{subsubsectionc}{0}\noindent 
        {\bf\thesectionc.\thesubsectionc. {\kern1pt \bfit #1}}\par\vspace{5pt}}
\renewcommand{\subsubsection}[1] {\vspace{12pt}\addtocounter{subsubsectionc}{1}
        \noindent{\tenrm\thesectionc.\thesubsectionc.\thesubsubsectionc.
        {\kern1pt \tenit #1}}\par\vspace{5pt}}
\newcommand{\nonumsection}[1] {\vspace{12pt}\noindent{\tenbf #1}
        \par\vspace{5pt}}

\newcounter{appendixc}
\newcounter{subappendixc}[appendixc]
\newcounter{subsubappendixc}[subappendixc]
\renewcommand{\thesubappendixc}{\Alph{appendixc}.\arabic{subappendixc}}
\renewcommand{\thesubsubappendixc}
        {\Alph{appendixc}.\arabic{subappendixc}.\arabic{subsubappendixc}}

\renewcommand{\appendix}[1] {\vspace{12pt}
        \refstepcounter{appendixc}
        \setcounter{figure}{0}
        \setcounter{table}{0}
        \setcounter{lemma}{0}
        \setcounter{theorem}{0}
        \setcounter{corollary}{0}
        \setcounter{definition}{0}
        \setcounter{equation}{0}
        \renewcommand{\thefigure}{\Alph{appendixc}.\arabic{figure}}
        \renewcommand{\thetable}{\Alph{appendixc}.\arabic{table}}
        \renewcommand{\theappendixc}{\Alph{appendixc}}
        \renewcommand{\thelemma}{\Alph{appendixc}.\arabic{lemma}}
        \renewcommand{\thetheorem}{\Alph{appendixc}.\arabic{theorem}}
        \renewcommand{\thedefinition}{\Alph{appendixc}.\arabic{definition}}
        \renewcommand{\thecorollary}{\Alph{appendixc}.\arabic{corollary}}
        \renewcommand{\theequation}{\Alph{appendixc}.\arabic{equation}}
        \noindent{\tenbf Appendix \theappendixc #1}\par\vspace{5pt}}
\newcommand{\subappendix}[1] {\vspace{12pt}
        \refstepcounter{subappendixc}
        \noindent{\bf Appendix \thesubappendixc. {\kern1pt \bfit #1}}
        \par\vspace{5pt}}
\newcommand{\subsubappendix}[1] {\vspace{12pt}
        \refstepcounter{subsubappendixc}
        \noindent{\rm Appendix \thesubsubappendixc. {\kern1pt \tenit #1}}
        \par\vspace{5pt}}

\newcommand{\textlineskip}{\baselineskip=13pt}
\newcommand{\smalllineskip}{\baselineskip=10pt}

\def\eightcirc{
\begin{picture}(0,0)
\put(4.4,1.8){\circle{6.5}}
\end{picture}}
\def\eightcopyright{\eightcirc\kern2.7pt\hbox{\eightrm c}} 

\newcommand{\copyrightheading}[1]
        {\vspace*{-2.5cm}\smalllineskip{\flushleft
        {\footnotesize International Journal of Modern Physics A, #1}\\
        {\footnotesize $\eightcopyright$\, World Scientific Publishing
         Company}\\
         }}

\def\abstracts#1#2#3{{
        \centering{\begin{minipage}{4.5in}\baselineskip=10pt\footnotesize
        \parindent=0pt #1\par 
        \parindent=15pt #2\par
        \parindent=15pt #3
        \end{minipage}}\par}} 

\renewenvironment{thebibliography}[1]
        {\frenchspacing
         \ninerm\baselineskip=11pt
         \begin{list}{\arabic{enumi}.}
        {\usecounter{enumi}\setlength{\parsep}{0pt}
         \setlength{\leftmargin 12.7pt}{\rightmargin 0pt} 
         \setlength{\itemsep}{0pt} \settowidth
        {\labelwidth}{#1.}\sloppy}}{\end{list}}

\newcounter{itemlistc}
\newcounter{romanlistc}
\newcounter{alphlistc}
\newcounter{arabiclistc}

\newcommand{\fcaption}[1]{
        \refstepcounter{figure}
        \setbox\@tempboxa = \hbox{\footnotesize Fig.~\thefigure. #1}
        \ifdim \wd\@tempboxa > 5in
           {\begin{center}
        \parbox{5in}{\footnotesize\smalllineskip Fig.~\thefigure. #1}
            \end{center}}
        \else
             {\begin{center}
             {\footnotesize Fig.~\thefigure. #1}
              \end{center}}
        \fi}

\newcommand{\tcaption}[1]{
        \refstepcounter{table}
        \setbox\@tempboxa = \hbox{\footnotesize Table~\thetable. #1}
        \ifdim \wd\@tempboxa > 5in
           {\begin{center}
        \parbox{5in}{\footnotesize\smalllineskip Table~\thetable. #1}
            \end{center}}
        \else
             {\begin{center}
             {\footnotesize Table~\thetable. #1}
              \end{center}}
        \fi}

\def\@citex[#1]#2{\if@filesw\immediate\write\@auxout
        {\string\citation{#2}}\fi
\def\@citea{}\@cite{\@for\@citeb:=#2\do
        {\@citea\def\@citea{,}\@ifundefined
        {b@\@citeb}{{\bf ?}\@warning
        {Citation `\@citeb' on page \thepage \space undefined}}
        {\csname b@\@citeb\endcsname}}}{#1}}

\newif\if@cghi
\def\cite{\@cghitrue\@ifnextchar [{\@tempswatrue
        \@citex}{\@tempswafalse\@citex[]}}
\def\citelow{\@cghifalse\@ifnextchar [{\@tempswatrue
        \@citex}{\@tempswafalse\@citex[]}}
\def\@cite#1#2{{$\null^{#1}$\if@tempswa\typeout
        {IJCGA warning: optional citation argument 
        ignored: `#2'} \fi}}

\def\pmb#1{\setbox0=\hbox{#1}
        \kern-.025em\copy0\kern-\wd0
        \kern.05em\copy0\kern-\wd0
        \kern-.025em\raise.0433em\box0}

\def\fnm#1{$^{\mbox{\scriptsize #1}}$}
\def\fnt#1#2{\footnotetext{\kern-.3em
        {$^{\mbox{\scriptsize #1}}$}{#2}}}

\def\fpage#1{\begingroup
\voffset=.3in
\thispagestyle{empty}\begin{table}[b]\centerline{\footnotesize #1}
        \end{table}\endgroup}

\font\tenrm=cmr10
\font\tenit=cmti10 
\font\tenbf=cmbx10
\font\bfit=cmbxti10 at 10pt
\font\ninerm=cmr9

\font\eightrm=cmr8

\def\qed{\hbox{${\vcenter{\vbox{                        
   \hrule height 0.4pt\hbox{\vrule width 0.4pt height 6pt
   \kern5pt\vrule width 0.4pt}\hrule height 0.4pt}}}$}}

 \def\unit{\hbox to 3.3pt{\hskip1.3pt \vrule height 7pt width .4pt \hskip.7pt
\vrule height 7.85pt width .4pt \kern-2.4pt
\hrulefill \kern-3pt
\raise 4pt\hbox{\char'40}}}

\begin{document}


\normalsize\textlineskip
\thispagestyle{empty}
\setcounter{page}{1}

\copyrightheading{}                     

\rightline{UG/00-15}
\rightline{VUB/TENA/00/07}

\vspace*{0.88truein}

\fpage{1}
\centerline{\bf TOWARDS A SUPERSYMMETRIC NON-ABELIAN}
\vspace*{0.035truein}
\centerline{\bf BORN-INFELD THEORY}
\vspace*{0.37truein}
\centerline{\footnotesize E. A. BERGSHOEFF, M. DE ROO}
\vspace*{0.015truein}
\centerline{\footnotesize\it Institute for Theoretical Physics,}
\baselineskip=10pt
\centerline{\footnotesize\it Nijenborgh 4, 9747 AG Groningen,}
\baselineskip=10pt
\centerline{\footnotesize\it The Netherlands}
\baselineskip=10pt
\centerline{\footnotesize\it 
E-mail: e.a.bergshoeff@phys.rug.nl, m.de.roo@phys.rug.nl}
\vspace*{10pt}
\centerline{\footnotesize A. SEVRIN}
\vspace*{0.015truein}
\centerline{\footnotesize\it Theoretische Natuurkunde,}
\baselineskip=10pt
\centerline{\footnotesize\it Vrije Universiteit Brussel,}
\baselineskip=10pt
\centerline{\footnotesize\it Pleinlaan 2, B-1050 Brussels, Belgium}
\baselineskip=10pt
\centerline{\footnotesize\it E-mail: asevrin@tena4.vub.ac.be}
\vspace*{0.225truein}

\vspace*{0.21truein}
\abstracts{We define an iterative procedure to obtain a
non-abelian generalization of the Born-Infeld
action. This construction is made possible by the
use of the severe restrictions imposed by kappa-symmetry.
We have calculated all bosonic terms in the action up to terms 
quartic in the Yang-Mills field strength and all fermion
bilinear terms up to terms cubic in the field strength.
Already at
this order the fermionic terms do not satisfy the
symmetric trace-prescription.}{}{}


\vspace*{1pt}\textlineskip      
\section{Abelian Born-Infeld}   
\vspace*{-0.5pt}
\noindent
One of the most intriguing features of D-branes is
their close connection to gauge theories.
Indeed, the effective theory describing the world-volume dynamics of a
D$p$-brane is a $p+1$--dimensional field theory with, in the static gauge,
as bosonic degrees of freedom the transversal coordinates $X^m\ (m = 1,
\cdots ,9-p)$ of the brane,
and the massless states of the open strings
ending on the brane which appear as a $U(1)$ gauge field describing $p-1$
degrees of freedom.
When these fields vary slowly, the effective
action governing their dynamics is known to all orders in $\alpha '$.
It is the ten-dimensional
Born-Infeld action \cite{AT}, dimensionally reduced to $p+1$ dimensions.
In the supersymmetric case there are additional fermionic degrees of freedom
$\chi_\alpha\ (\alpha = 1,\cdots 16)$ describing 8 fermionic degrees
of freedom so that we have a total of $8+8$ degrees of freedom.

Clearly, the $9-p$ transversal scalars break the 10-dimensional
Lorentz covariance.
The fully covariant worldvolume theory of a single D-brane in a type
II theory can
be formulated in terms of the following worldvolume fields \fnm{a}\fnt{a}{
The indices $\mu,\nu=0,\ldots,9$ are spacetime indices whereas the indices
$i,j=0,\ldots,p$ label the worldvolume coordinates $\sigma^i$.}: the
embedding coordinates $X^\mu(\sigma)$ (of which only the transverse
coordinates represent physical degrees of freedom), the Born-Infeld vector
field $V_i(\sigma)$, and $N=2$ spacetime fermionic fields
$\theta(\sigma)$.
The presence of a local fermionic symmetry ($\kappa$-symmetry)
makes it possible to gauge away half the fermionic
degrees of freedom. The field content then corresponds, in a static gauge,
to that of a supersymmetric Yang-Mills theory in $p+1$ dimensions
describing $8+8$ degrees of freedom.

In this talk our main emphasis will be on the IIB D9--brane. This case
is special in the sense that, for $p=9$, there are no transverse scalars
and the  worldvolume theory is given by a D=10 supersymmetric
Maxwell multiplet $(V_\mu, \chi)$. The bosonic part of the 
action is given by the ten-dimensional Born-Infeld Lagrangian 
\cite{Fradkin:1985qd,Abouelsaood:1987gd}

\begin{equation}
{\cal L}({\rm bosonic}) = - \sqrt {- {\rm det}\,\big (
\eta_{\mu\nu} + 2\pi \alpha^\prime F_{\mu\nu}\big )}\,\, .
\end{equation}
Expanding the square root this action gives rise to an infinite number of 
terms with even powers of $F$. The term quadratic in $F$ is the usual
Maxwell kinetic term. 

In the supersymmetric case every $F^n$ term gets a supersymmetric partner
and the complete action has the schematic form

\begin{eqnarray}
{\cal L}({\rm supersymmetric}) &=&  {\rm tr} F^2
+ (2\pi \alpha^{\prime})^{2} 
\big [ {\rm tr} F^4 - {\textstyle{1\over 4}}\big (
{\rm tr} F^2\big )^2\big ] + \cdots\nonumber\cr
&&\cr
&+& \bar\chi \partial \chi\ + (2\pi \alpha^{\prime})^{2} \, F^2
\bar\chi \partial \chi 
+ \cdots
\end{eqnarray}
The fermionic $F^2\bar\chi\partial\chi$ terms were found in 
\cite{Bergshoeff:1987jm,Metsaev:1987by}
from the requirement that the action must 
possess a linear supersymmetry. 
Complete results were obtained in D=4
dimensions via superspace techniques \cite{Cecotti:1987gb}.

Recently a new development took place which 
led to a construction of the complete supersymmetric Born-Infeld
action. A crucial role in this development was played by
kappa-symmetry.
The kappa-symmetric covariant D-brane actions were constructed
in flat  \cite{Aga}, as well as in curved backgrounds
\cite{Ced,BT}.

Schematically, the kappa-symmetric Lagrangean is given by \fnm{b}\fnt{b}{
From now on we set $2\pi\alpha^{\prime} = 1$.}

\begin{equation}\label{BIA}
{\cal L}_{\rm BI} = - e^{-\phi} \sqrt {-{\rm det}\,
\big (g+{\cal F}\big )}\ +\ Ce^{\cal F}\, ,
\end{equation}
with ${\cal F} = 2dV +B$.
The first term is often called the kinetic term whereas the second one
is the Wess-Zumino term.
It is understood that all NS-NS background fields $g,\phi, B$ and
all R-R background fields $C$ are superfields defined over a superspace
with coordinates $(X^\mu,\theta)$. 
The local $\kappa$-symmetry acts on the fermions as
\begin{equation}
\label{dt}
 \delta\bar\theta(\sigma) = \bar \eta (\sigma) \equiv
\bar\kappa(\sigma)\,(1+\Gamma)\,,
\end{equation}
where $\Gamma$, which depends on worldvolume as well as background fields,
satisfies
\begin{equation}
  \Gamma^2 = \unit \,.
\end{equation}
The projection (\ref{dt})
makes it possible to gauge away half the fermionic
degrees of freedom.
The variation of the D-brane action takes the form
\begin{equation}
\label{varkappa}
  \delta{\cal L} = - \bar\eta\, (1-\Gamma) {\cal T}\,,
\end{equation}
where $\cal T$ is some expression in terms of the worldvolume and
background fields.
The variation (\ref{varkappa})
indeed vanishes if $\bar\eta$ is given by (\ref{dt}).
This variation has the following source: the first 
term within the round
brackets comes from the kinetic  term in (\ref{BIA}),
the term with $\Gamma$ arises from the Wess-Zumino contribution.

From the above it is clear that a crucial role is played by the
projection operator $1 + \Gamma(X,{\cal F})$. In the case
of the IIB D9--brane the ${\cal F}$-independent part $\Gamma^{(0)}(X)$
of $\Gamma(X,{\cal F})$ is given by

\begin{equation}
\label{gamma0}
  \Gamma^{(0)}(X) = {1\over 10!\sqrt{-\det g}}
   \epsilon^{i_1\ldots i_{10}}\gamma_{i_1\ldots i_{10}}\, ,
\hskip 1truecm  \left(\Gamma^{(0)}(X) \right)^2 = \unit\,.
\end{equation}
The worldvolume metric reads
\begin{equation}
  g_{ij} = \eta_{ij} + \bar\theta \gamma_{(i}\partial_{j)}\theta\,,\qquad
  \eta_{ij} \equiv \partial_i X^\mu \partial_j X_\mu\,,
\end{equation}
and
\begin{equation}
   \gamma_i \equiv \Gamma^\mu\partial_i X^\mu\,.
\end{equation}

The complete expression for $\Gamma(X,{\cal F})$ is given 
by \cite{BT}

\begin{equation}
\label{fullgamma}
  \Gamma = {\sqrt{-\det g}\over \sqrt{-\det(g+{\cal F})}}
     \Gamma^{(0)}
     \sum_{k=0}^5 {1\over 2^k k!} {\cal P}_{(k)} \gamma^{i_1\ldots i_{2k}}
       {\cal F}^{k}_{i_1\ldots i_2k}\, ,
\end{equation}
where
\begin{equation}
  {\cal P}_{(k)} =  \sigma_1\ ({\rm for}\ k=1,3,5)\,,
           \quad {\cal P}_{(k)} = i\sigma_2\ ({\rm for}\ k=2,4)\,.
\end{equation}
The Pauli-matrices refer to the fact that we are using an N=2
superspace notation.

It was realized in \cite{Aga}
that for the special case of a IIB D9--brane in a
flat background, the results take on a particularly simple form.
We consider a flat background, i.e.~we take $e_\mu{}^a = \delta_\mu{}^a$
and all
other fields zero. Furthermore, we fix the kappa-transformations and
worldvolume parametrizations (we choose the static gauge) by setting

\begin{equation}
X^\mu = \delta^{i\mu}\sigma^i\, ,
\hskip 1.5truecm \theta = \pmatrix{\theta_1\equiv \chi\cr
                                   \theta_2 = 0}\, .
\end{equation}
The result of this gauge-fixing is that the complete Wess-Zumino term
vanishes and the kinetic term
is given by the simple expression \cite{Aga}

\begin{equation}
{\cal L}_{\rm BI} = - \sqrt {- {\rm det}\, \big (
\eta_{\mu\nu} + F_{\mu\nu} + \bar\chi\Gamma_\mu\partial_\nu\chi +
{\textstyle{1\over 4}}\bar\chi \Gamma^a\partial_\mu\chi \bar\chi\Gamma_a
\partial_\nu\chi\big )}\,\, .
\end{equation}
This action has 16 linear supersymmetries as well as 16 nonlinear
 supersymmetries. The explicit form of these supersymmetry rules  can be 
obtained from the original 32 supersymmetries of the N=2 superspace
after taking care of the fact that they get deformed with
a field-dependent kappa-transformation upon fixing the kappa-gauge.

Our goals, which we will discuss in this talk, are two-fold:

\begin{description}
\item{(1)} We want to find the non-abelian version of the projection
operator $1+\Gamma$. One application we have in mind is that
knowing the explicit expression of $\Gamma$ enables us to investigate
non-abelian BPS configurations. The condition of supersymmetry
for such configurations is given by $(1-\Gamma)\epsilon = 0$
\cite{BKOP}.

\item{(2)}
A second, and more ambitious goal is to construct the non-abelian 
supersymmetric Born-Infeld action by the use of the
severe restrictions of kappa-symmetry as suggested in
\cite{Denef:2000rj}. We restrict ourselves to a flat background
for reasons to be discussed later (see the comments section). Non-abelian
N=2, D=4 supersymmetric Born-Infeld
actions (with a symmetrized trace condition) have recently been
constructed in \cite{Ketov:2000fv,Refolli:2000fv}.

\end{description}

It should be stressed that we only discuss the IIB D9--brane. 
We expect that at least some of the structure of the lower-dimensional
branes can be obtained by applying T-duality
\cite{Taylor:2000pr,Myers:1999ps}. These lower-dimensional cases 
have the additional complication that they contain non-trivial
transverse scalars. It is an open question what kind of ``D-geometry''
these scalars are supposed to describe \cite{Douglas:2000vm,deboer}.

\section{Non-Abelian Born-Infeld}
\noindent
In the case of a single D--brane the worldvolume theory, in lowest
order in $\alpha^\prime$ is given by a supersymmetric Maxwell theory.
The result to all orders in $\alpha^\prime$ is given by the supersymmetric
BI action.

Once several D-branes are present, the situation changes. The mass of the
strings stretching between two branes is proportional to the shortest
distance between the two branes. Starting off with $n$ well separated
D-branes, we end up with a $U(1)^n$ theory. However, once the $n$ branes
coincide, additional massless states appear which complete the gauge
multiplet to a non-abelian $U(n)$-theory \cite{EW1}. 
Contrary to the abelian case,
the effective action is not known to all orders in $\alpha '$. The first
term, quadratic in the field strength, is nothing but a dimensionally
reduced $U(n)$ Yang-Mills theory. The next order, which is quartic in the
field strength, was obtained from the four-gluon scattering amplitude
in open superstring theory \cite{GW} and
from a three-loop $\beta$-function calculation \cite{BP}.

The notion of an
effective action for slowly varying fields is subtle in the non-abelian
case \cite{AT}. In the effective action  higher derivative terms are dropped.
However because of
\begin{eqnarray}
D_i D_j F_{kl}=\frac 1 2 \{D_i , D_j\}F_{kl}-\frac i 2 {[}F_{ij}, F_{kl}
{]},
\end{eqnarray}
this is ambiguous. The analysis of the mass spectrum seems to indicate
that the symmetrized product of
derivatives acting on a fieldstrength should be viewed as an acceleration
term which can safely be neglected, while the anti-symmetrized products
should be kept.

Different order prescriptions of the non-abelian matrices occurring
in the non-abelian Born-Infeld action have been given in the literature.
In \cite{AT2}  a symmetrized trace prescription was suggested: the non-abelian
Born-Infeld action assumes essentially the same form as the abelian one,
however, all Lie algebra valued objects have to be symmetrized first before
taking the trace. Other trace prescriptions, involving
commutators, have been given as well
\cite{Argyres:1990qr}.

Recently, it was found that the symmetrized trace prescription could
not be correct as it did not reproduce the mass spectrum of certain
D-brane configurations \cite{HT,Denef:2000rj}.
It was shown in \cite{B}
that by adding commutator terms to the action the problem
might be cured.

For $n$ overlapping D9-branes the completely gauge-fixed result
should be the supersymmetric version of the non-abelian Born-Infeld
theory. Since the vector fields $V_i^A(\sigma)$, $A=1,\ldots,n^2$,
are in the adjoint representation of $U(n)$, we have to
make the same choice for the fermion fields $\theta$.
Therefore we start out with fields $\theta^A(\sigma)$,
which form a doublet
$(N=2)$ of Majorana-Weyl spinors for each $A$,
satisfying $\Gamma_{11}\theta^A=\theta^A$. After $\kappa$-gauge-fixing
only half of each doublet will remain, and we have the
correct number of degrees of freedom for the supersymmetric Yang-Mills
theory.

This requires, that there are as many $\kappa$-symmetries as $\theta$'s,
so that also the parameter of the $\kappa$ transformations will
have to be in the adjoint of $U(n)$. Thus the $\theta^A$
transform as follows under
ordinary supersymmetry ($\epsilon$), $\kappa$-symmetry ($\kappa$),
Yang-Mills transformations $(\Lambda^A)$, and worldvolume
reparametrisations ($\xi^i$):
\begin{equation}
\label{dtall}
   \delta\bar\theta^A(\sigma) = -\bar\epsilon^A
     + \bar\eta^A(\sigma)
     + f^A{}_{BC}\Lambda^B(\sigma)\bar\theta^C(\sigma)
     + \xi^i(\sigma)\partial_i \bar\theta^A(\sigma)\,,
\end{equation}
with $\bar\eta^A \equiv  \bar\kappa^B(\unit\,\delta^{BA} +
        \Gamma^{BA})$.
Here $\epsilon^A$ are constant, $\Gamma^{AB}$ depends on the
worldvolume fields, and therefore on $\sigma$. It must satisfy
\begin{equation}
  \Gamma^{AB}\Gamma^{BC} = \delta^{AC}\unit\,.
\end{equation}

Useful information is obtained by considering commutators of
these transformations.
Because $\epsilon^A$ is constant we find from the commutator of
Yang-Mills and supersymmetry transformations that
\begin{equation}
      f^{A}{}_{BC}\Lambda^B\epsilon^C=0\to f_{ABC}\epsilon^C=0\,.
\end{equation}
Therefore $\epsilon = \epsilon^A T_A$, where $T_A$ are the
$U(n)$ generators, must be proportional to the unit
matrix, i.e.~we can choose a basis in which there is only
one nonvanishing $\epsilon$ parameter.
So only a subset of the $\theta^A$ transform under supersymmetry,
and there is only one independent supersymmetry parameter.
The $\theta$'s which are presently
inert under supersymmetry will obtain their supersymmetry
transformations after $\kappa$-gauge fixing.

The only spacetime scalars we have are  the embedding coordinates
$X^\mu(\sigma)$ for worldvolume directions.
There are several options that one could consider for the $X^\mu$:
\begin{enumerate}
\item We could assume that we are in the static gauge, i.e.
\begin{equation}
    X^\mu(\sigma) = \delta^\mu_i\,\sigma^i\,,
\end{equation}
 from the
 beginning, so that the $X^\mu$ are absent. In this case there are no
 worldvolume reparametrisations, i.e.~$\xi^i=0$ in (\ref{dtall}).
\item We could decide that the $X^\mu$ are in the singlet representation
 of the Yang-Mills group. The idea is that the $n$ branes overlap,
 there is only one set of worldvolume coordinates, and the corresponding
 reparametrization group would be sufficient to gaugefix a singlet
 set of embedding coordinates.
\item We could choose the $X^\mu$ in the adjoint representation of
 Yang-Mills in analogy with transverse coordinates for $p<9$.
 Here one thinks of starting with $n$ separate branes where each has its
 own worldvolume and embedding coordinates. When the branes overlap
 the embedding coordinates "fill up" to form elements of the adjoint
 representation.
 Clearly
 this requires a different approach towards the world-volume
 reparametrisation invariance, which must then correspond to a sufficiently
 large symmetrygroup to gaugefix all these embedding functions.
\end{enumerate}
We have investigated the first two possibilities in the non-abelian case,
and we have found that only the first approach is consistent with the
iterative procedure that we employ.

Our strategy will be to construct a kappa-symmetric action via an
iterative procedure. 
We found that the flat background expressions of the abelian NS-NS and R-R
fields cannot be generalized to the non-abelian case while remaining
YM singlets. We therefore decided to make the most general Ansatz
for the action and to require kappa-symmetry, order
by order in $F$ and up to terms quartic in the fermions.
The iteration is obtained by expanding $\Gamma$ and ${\cal T}$ 
in the variation (\ref{varkappa}) order by order in $F$:
\begin{eqnarray}
  \delta{\cal L} &=& -\bar\eta\,(1-(\Gamma_0+\Gamma_1+\ldots))
       ({\cal T}_0 + {\cal T}_1  +\ldots ) \nonumber\\
       &=& - \bar\eta\, ({\cal T}_0 - \Gamma_0{\cal T}_0
         + {\cal T}_1 - \Gamma_1{\cal T}_0 - \Gamma_0{\cal T}_1  +\ldots)\,,
\end{eqnarray}
where the subindices reflect the order in $F$.
We completed this procedure up to variations of the action quadratic in $F$.
Due to lack of space we give the result, 
both in its kappa-symmetric as well as in its gauge-fixed form,
not here, but in a forth-coming publication \cite{us}.

\section{Comments}
\noindent
We have been able to embed a D=10 supersymmetric $U(n)$ YM theory
(up to terms in the action quartic in $F$) into a kappa-symmetric system.
In particular, we find fermionic terms of the form
$[F^2,\bar\chi]\,\partial \chi$ which violate the symmetrized trace prescription.

We hope that our partial results will contribute to a better understanding
of the non-abelian Born-Infeld action and perhaps may lead to
the complete answer like in the abelian case. In this respect it is
of interest to remember that the abelian expression (\ref{fullgamma})
for $\Gamma(X,F)$ can be written as \cite{BKOP}

\begin{equation}
\Gamma(X,F) = e^{-a/2} \Gamma^{(0)}(X) e^{a/2}\, ,
\end{equation}
with 

\begin{equation}
a = {\textstyle{1\over 2}}Y_{ij}\Gamma^{ij}\sigma_3\, .
\end{equation}
The matrix $Y$ is related to the matrix $F$ by a so-called
``tan'' relation defined in \cite{BKOP}: $F =$ ``tan''$\, Y$. 
These results were
obtained by looking to branes at angles 
\cite{Berkooz:1996km,Balasubramanian:1997uc}.
Via T-duality these systems are related to branes with $F\ne 0$
as we discuss in this talk. It would be interesting to see whether a
similar simple expression can be found for the nonabelian case

Another approach to find the complete answer could be to use the superembedding
techniques developed in \cite{Bandos:1995zw,Howe:1997wf}.
Finally, yet another way to get the complete answer could be
to extend to the non-abelian case the analysis of \cite{Bellucci:2000bd}
where it was shown how the superworldvolume dynamics of 
superbranes can be obtained form nonlinear realizations.

We would like to end by mentioning a few important open issues.

\begin{description}

\item{(A)} Can the results we obtained be generalized to a supersymmetric
curved background? On the one hand the flat background expressions
we find do not fit into general supergravity background field expressions
that are YM singlets. We do not allow the supergravity fields
to be in a nontrivial representations of the YM group.
One possible scenario could be that the D-brane only couples
to the U(1) part of the N=1 supergravity background fields and that
the off-diagonal terms only involve worldvolume fields. The price
to pay is that not all worldvolume fields are on the same footing 
whereas, at least after fixing the kappa-symmetries, they should be
\fnm{c}\fnt{c}{We thank A.~Tseytlin for a discussion on this issue.}.

\item{(B)} A somewhat related issue is: is there a natural
(non-abelian) superspace geometry that describes the kappa-symmetric
result? It is well-known that, to describe the abelian
kappa-symmetric action, an important role is played by 
ordinary superspace
geometry. In the non-abelian case
we are dealing with many fermionic coordinates on which just 
a single N=2 supersymmetry is realized. Usually, introducing more fermionic
coordinates, means going to an extended superspace with extended
supersymmetries but this is not the case in the present situation.

\item{(C)} It would be interesting to see what happens
when we T-dualize our results and obtain the worldvolume
theory of non-abelian Dp--branes with $p<9$. Our hope is that
this will teach us something about the D-geometry which is involved
in describing such systems.

\item{(D)} Finally, our results can be used to search for 
non-Abelian BPS configurations, i.e.~configurations with 
nontrivial $F$-terms that take values outside the Cartan subalgebra.

\end{description}

\nonumsection{Acknowledgements}
\noindent
One of us (E.B.) would like to thank the organizers of Strings 2000 for
providing such a stimulating environment. We like to thank 
I.~Bandos,
M.~Cederwall,
S.~Ferrara,
S.F.~Hassan,
E.~Ivanov,
R.~Kallosh,
U.~Lindstr\"om,
C.~Nappi,
A.~Peet,
V.~Periwal,
D.~Sorokin,
J.~Troost and
A.~Tseytlin.
for useful discussions. This work is supported by the European Commission
RTN programme HPRN-CT-2000-00131, in which E.B. and M.d.R. are associated 
to the university of Utrecht and A.S. is associated to the university of 
Leuven.

\end{document}